\documentclass[prb,showpacs]{revtex4}
\usepackage{amsfonts}
\usepackage{amssymb}
\usepackage{graphicx}
\usepackage{dcolumn}
\usepackage{amsmath}

\begin{document}

\title{Boundaries of Subcritical Coulomb Impurity Region in Gapped Graphene }
\author{B. S. Kandemir and A. Mogulkoc}
\affiliation{Department of Physics, Faculty of Sciences, Ankara University, 06100\\
Tando\u{g}an, Ankara, Turkey}
\date{\today}

\begin{abstract}
The electronic energy spectrum of graphene electron subjected to a
homogeneous magnetic field in the presence of a charged Coulomb impurity is
studied analytically within two-dimensional Dirac-Weyl picture by using
variational approach. The variational scheme we used is just based on
utilizing the exact eigenstates of two-dimensional Dirac fermion in the
presence of a uniform magnetic field as a basis for determining analytical
energy eigenvalues in the presence of an attractive/repulsive charged
Coulomb impurity. This approach allows us to determine under which
conditions bound state solutions can or can not exist in gapped graphene in
the presence of magnetic field. In addition, the effects of uniform magnetic
field on the boundaries of subcritical Coulomb impurity region in the
massless limit are also analyzed. Our analytical results show that the
critical impurity strength decreases with increasing gap/mass parameter, and
also that it increases with increasing magnetic field strength. In the
massless limit, we investigate that the critical Coulomb coupling strength
is independent of magnetic field, and its upper value for the ground-state
energy is $0.752$.
\end{abstract}

\maketitle

Since the discovery of graphene\cite{novoselov} , strictly two-dimensional
(2D) carbon system with hexagonal lattice structure, a great deal of both
experimental and theoretical research efforts has been achieved to identify
the electronic structures of graphene and graphene based nanostructures such
as graphene dots and graphene nanoribbons\cite{general}. In fact, from the
theoretical point of view, after Wallace's work\cite{wallace}, Semenoff\cite%
{semenoff} was the first to construct the relativistic 2+1 dimensional
electrodynamics analogue of tight-binding model of electrons in a 2D
hexagonal lattice. In other words, he investigated that, in the long-wave
length (continuum) limit, the low-energy electronic structure of this model
exhibits linearly dispersing massless fermionic quasi particles around the
degeneracy points obeying to 2+1 Dirac-Weyl Hamiltonian. As a consequence,
all the single particle low-energy physics of graphene is believed to be
well simulated by the Dirac-Weyl equation.

Due to their relevance for the transport properties, impurity effects are,
at present, at the center of both theoretical and experimental
investigations in graphene physics \cite%
{mirlin,kats2,valeri,fogler,kotov,bruno,wei,impurity}. There are
non-negligible effects especially on the electronic properties of graphene
based devices, since the Dirac points of undoped graphene which are very
sensitive to such effects\cite{vincenzo}. Additionally, a recent experiment%
\cite{chen} performed by depositing potassium atoms onto undoped graphene
addresses charged Coulomb like behavior. Moreover, it is shown that a
pronounced asymmetry occurs in transport cross-section depending upon $%
Z\rightarrow -Z$ for the Coulomb scattering\cite%
{shytov,shytov2,novikov,novikov2}, and it is also believed that the
long-range charged Coulomb impurity limit the mobility of graphene. \cite%
{chen,kats}.

From the quantum electrodynamical (QED) point of view, it is well-known that
the interactions between the charged particles are associated by the
exchange of virtual photons. These interactions are represented by a
fermionic Coulomb potential which is identified with its perturbative
expansion in powers of the fine structure $\alpha =e^{2}/\hbar c=1/137$.
However, it is broken down in the case of superheavy nuclei, atoms or
quasi-molecules where $\alpha $ is of order of unity \cite{greiner}. This
hypothetical regime with $Z\alpha >1$ which requires a non-perturbative
treatment (due to difficulties of convergence of expansions) has been of
academic interest in the past. However, due to the fact that the effective
coupling constant of graphene is at order of $\alpha =e^{2}/\varepsilon
\hbar v_{F}\sim 1$ for conventional SiO$_{2}$ substrate\cite%
{kats2,valeri,fogler,shytov,shytov2},  condensed matter analog of
this regime is of experimental interest, and it is now at the center
of theoretical investigations. In particular, "mass" or energy gap
in graphene is frequently interpreted as being related to chiral
symmetry breaking of 2D massless Dirac-Weyl fermions induced by
substrate\cite{zhou}, spin-orbit coupling\cite{kane}, and by
boundary conditions\cite{Nakada}. Since this enables one to tune a
gap by just adjusting the external parameters of graphene, it is of
particular interest itself.

In the present paper, we investigate analytically the effects of a Coulomb
impurity with effective charge $Ze$ onto the energy spectrum of 2D massive
fermions, i.e., gapped graphene, with a uniform magnetic field perpendicular
to the graphene plane. In other words, we analyze the effects of uniform
magnetic field onto the graphene electron with impurity, and hence we
determine the boundaries of subcritical regime. As mentioned above, due to
both strong coupling and long-range characters of the Coulomb impurities in
graphene, these effects can not be addressed by conventional perturbative
techniques. Therefore, we suggest a variational approach to account for such
effects, induced by the impurity.

The Dirac Hamiltonian with electromagnetic potentials that we
describe the motion of an electron and a hole in 2D graphene
subjected to a constant uniform magnetic field, perpendicular to
graphene plane, can be written in
the form%
\begin{equation}
H=\hbar v_{F}\ \boldsymbol{\alpha }\cdot \left( \mathbf{-}i\partial _{\mu }%
\mathbf{+}\frac{e}{\hbar c}\mathbf{A}\right) +\boldsymbol{\beta }%
mv_{F}^{2}-eA_{0}(r),  \label{1}
\end{equation}%
where $A_{0}$ and $A_{\mu }$ are the time and space components of the
four-vector potential $\left( A_{0},A_{\mu }\right) ,$ due to impurity $%
A_{0}=Ze/\varepsilon r$ and static uniform magnetic field with $\mathbf{A}%
=B\left( -y,x\right) /2$, respectively. Here, while $Z>0$ refers to the
attractive impurity potential that binds electrons to impurity and repels
holes, or vice versa if $Z<0$. In Eq.~(\ref{1}), we have used the
Dirac-Pauli representation of Dirac matrices, $\boldsymbol{\alpha }$ and $%
\boldsymbol{\beta }$, each of which are written in two by two-block form.
Using two component spinor representation as $\Psi ^{\dag }\boldsymbol{=}$ $%
\begin{pmatrix}
\phi ^{\ast } & \chi ^{\ast }%
\end{pmatrix}%
$, we see that each component of the eigenvalue equation $H\Psi =E\Psi $
satisfies the following coupled first order equation:%
\begin{eqnarray}
\boldsymbol{\sigma }\mathbf{\cdot }\left( \mathbf{-}i\partial _{\mu }\mathbf{%
+}\frac{e}{\hbar c}A_{\mu }\right) \chi +\left[ M_{0}-\overline{E}-e%
\overline{A}_{0}(r)\right] \phi  &=&0  \notag \\
\boldsymbol{\sigma }\mathbf{\cdot }\left( \mathbf{-}i\partial _{\mu }\mathbf{%
+}\frac{e}{\hbar c}A_{\mu }\right) \phi -\left[ M_{0}+\overline{E}+e%
\overline{A}_{0}(r)\right] \chi  &=&0,  \label{2}
\end{eqnarray}%
where we write energy, coupling strength and "mass" terms in units of $\hbar
v_{F}$ , as $\overline{E}=E/\hbar v_{F}$, $\overline{A}_{0}(r)=Ze/\hbar
v_{F}r$, and $M_{0}=mv_{F}/\hbar $, respectively. From our knowledge on
planar relativistic and non-relativistic \ electron systems in the presence
of both Coulomb and magnetic field, it is not possible to find exact
analytical solutions\cite{x1,x2} of Eq.~(\ref{2}) due to having hidden $%
s\ell _{2}$ algebraic structure\cite{x3}.

Decoupling Eq.~(\ref{2}) in the absence of impurity potential one can easily
see that the upper component of $\Psi $, i.e., $\phi ,$ should satisfy the
second order equation

\begin{equation}
\left[ -\frac{\partial ^{2}}{\partial \rho ^{2}}-\frac{1}{\rho }\frac{%
\partial }{\partial \rho }-\frac{1}{\rho ^{2}}\frac{\partial ^{2}}{\partial
\varphi ^{2}}-i\frac{eB}{\hbar c}\frac{\partial }{\partial \varphi }+\frac{eB%
}{\hbar c}\sigma _{3}+\left( \frac{eB}{2\hbar c}\right) ^{2}\rho ^{2}+\left( M_{0}^{2}-%
\overline{E}^{2}\right) \right] \phi =0,  \label{3}
\end{equation}%
whose solutions can then be written in terms of Laguerre
polynomials\cite{bhatta}.

\begin{equation}
\phi _{\nu m,s}(\rho ,\varphi )=\left( \sqrt{\frac{eB}{2\hbar c}}\rho
\right) ^{\left\vert m\right\vert }e^{im\varphi }e^{-eB\rho ^{2}/4\hbar
c}L_{\nu }^{\left\vert m\right\vert }\left( \frac{eB}{2\hbar c}\rho
^{2}\right) ,  \label{4}
\end{equation}%
provided that the energy quantization condition

\begin{equation}
\left[ 4\nu +2\left( \left\vert m\right\vert +1\right) +2\left( s+m\right) %
\right] \frac{eB}{2\hbar c}=\overline{E}^{2}-M_{0}^{2},  \label{5}
\end{equation}%
is fulfilled. It is obvious that Eq.~(\ref{3}) yields infinitely degenerate
energy eigenvalues $\overline{E}_{n}^{\pm }=\pm \left[ M_{0}^{2}+\left(
2neB/\hbar c\right) \right] ^{1/2}$ with $n=\nu +\left[ \left( \left\vert
m\right\vert +m+s+1\right) /2\right] .$ Inserting Eq.~(\ref{4}) into Eq.~(%
\ref{2}), the other component of the spinor can explicitly be determined.
Therefore, to the positive energy spinors in the absence of impurity
potential, we can write the complete and orthonormalized solutions of Eq.~(%
\ref{1}) for $s=+1$ and $s=-1$ cases as

\begin{equation}
\Psi _{\nu m,+1}^{+}(\rho ,\varphi )=\frac{1}{\sqrt{\pi }}\left[ N_{\nu
m}^{+}\left( M_{0},B\right) \right] ^{1/2}\gamma ^{\left\vert m\right\vert
+1}\rho ^{\left\vert m\right\vert }\exp \left( im\varphi \right) \exp \left(
-\gamma ^{2}\rho ^{2}/2\right)
\begin{bmatrix}
\mathbb{L}_{\nu m,+1}^{1+}\left( \gamma \rho \right)  \\
0 \\
0 \\
\mathbb{L}_{\nu m,+1}^{2+}\left( \gamma \rho \right)
\end{bmatrix}
\label{6}
\end{equation}%
and%
\begin{equation}
\Psi _{\nu m,-1}^{+}(\rho ,\varphi )=\frac{1}{\sqrt{\pi }}\left[ N_{\nu
m}^{+}\left( M_{0},B\right) \right] ^{1/2}\gamma ^{\left\vert m\right\vert
+1}\rho ^{\left\vert m\right\vert }\exp \left( im\varphi \right) \exp \left(
-\gamma ^{2}\rho ^{2}/2\right)
\begin{bmatrix}
0 \\
\mathbb{L}_{\nu m,-1}^{1+}\left( \gamma \rho \right)  \\
\mathbb{L}_{\nu m,-1}^{2+}\left( \gamma \rho \right)  \\
0%
\end{bmatrix}%
,  \label{7}
\end{equation}%
respectively, with the abbreviations%
\begin{eqnarray*}
\mathbb{L}_{\nu m,+1}^{2+}\left( \gamma \rho \right)  &=&\frac{i\gamma \rho
\exp (+i\varphi )}{\Gamma \left( M_{0},n\right) }\left\{
\begin{array}{lr}
L_{\nu }^{\left\vert m\right\vert +1}\left( \gamma ^{2}\rho ^{2}\right)  &
\text{\ }m\geq 0 \\
-\frac{\nu +1}{\gamma ^{2}\rho ^{2}}L_{\nu +1}^{\left\vert m\right\vert
-1}\left( \gamma ^{2}\rho ^{2}\right)  & \text{\ \ }m<0%
\end{array}%
\right. \ \ \ \ \  \\
\ \ \ \ \ \mathbb{L}_{\nu m,-1}^{2+}\left( \gamma \rho \right)  &=&\frac{%
i\gamma \rho \exp (-i\varphi )}{\Gamma \left( M_{0},n\right) }\left\{
\begin{array}{lr}
-\frac{\nu +\left\vert m\right\vert }{\gamma ^{2}\rho ^{2}}L_{\nu
}^{\left\vert m\right\vert -1}\left( \gamma ^{2}\rho ^{2}\right)  & \text{\ }%
m\geq 0 \\
L_{\nu -1}^{\left\vert m\right\vert +1}\left( \gamma ^{2}\rho ^{2}\right)  &
\text{\ \ }m<0%
\end{array}%
\right. \ \ ,
\end{eqnarray*}%
and $\mathbb{L}_{\nu m,+1}^{1+}\left( \gamma \rho \right) =\mathbb{L}_{\nu
m,-1}^{1+}\left( \gamma \rho \right) =L_{\nu }^{\left\vert m\right\vert
}\left( \gamma ^{2}\rho ^{2}\right) $, where $L_{\nu }^{\left\vert
m\right\vert }$ are the well-known associated Laguerre polynomials . In Eq.~(%
\ref{6}) and Eq.~(\ref{7}), we have also defined
\begin{equation*}
N_{\nu m}^{+}\left( M_{0},B\right) =\frac{\nu !}{(\nu +\left\vert
m\right\vert )!\ }\frac{M_{0}+\left[ M_{0}^{2}+\left( 2n/\ell ^{2}\right) %
\right] ^{1/2}}{2\left[ M_{0}^{2}+\left( 2n/\ell ^{2}\right) \right] ^{1/2}},
\end{equation*}%
and%
\begin{equation*}
\Gamma \left( M_{0},n\right) =\frac{\ell}{\sqrt{2}}\left\{
M_{0}+\left[ M_{0}^{2}+\left( 2n/\ell ^{2}\right) \right]
^{1/2}\right\} ,
\end{equation*}%
where $\gamma ^{2}=eB/2\hbar c$, and $\ell $ is the magnetic
confinement length  given by $\ell =\sqrt{\hbar c/eB}$. Analogously,
one can follow the same treatment for the negative energy spinors.

At this work, we suggest the states given by Eq.~(\ref{6}) and Eq.~(\ref{7})
as the trial states for the whole system, i.e., 2D Dirac equation with
charged Coulomb impurity given by Eq.~(\ref{2}). Therefore, for the
expectation value of H given by Eq.~(\ref{1}), we obtain

\begin{equation}
\overline{E}_{nms}^{\mp }\left( \gamma \right) =\int d^{2}\mathbf{r}\Psi
_{\nu m,\mp 1}^{\mp \ \dag }\left( \gamma \right) \left[ \boldsymbol{\alpha }%
\mathbf{\cdot }\left( \mathbf{-}i\partial _{\mu }\mathbf{+}\frac{e}{\hbar c}%
A_{\mu }\right) +\boldsymbol{\beta }M_{0}-\widetilde{A}_{0}(r)\right] \Psi
_{\nu m,\mp 1}^{\mp }\left( \gamma \right)   \label{8}
\end{equation}%
where $\widetilde{A}_{0}(r)=Z\alpha /r$, and $\alpha $ is given by $\alpha
=e^{2}/\varepsilon \hbar v_{F}$. By using first Eq.~(\ref{6}) together with
Eq.~(\ref{7}) in Eq.~(\ref{8}), and then performing the necessary integrals
\cite{grad}, the total variational energy $\overline{E}_{nms}^{\pm }\left(
\gamma \right) $ of the whole system can finally be written in the form

\begin{equation}
\overline{E}_{nms}^{\mp }\left( \gamma \right) =\left[ 1+\frac{n}{\Gamma
^{2}\left( M_{0},n\right) }\right] ^{-1}\left\{ M_{0}\left[ 1-\frac{n}{%
\Gamma ^{2}\left( M_{0},n\right) }\right] +\left[ \frac{2n}{\Gamma \left(
M_{0},n\right) }-Z\alpha \mathbb{M}_{\nu m,s}^{\mp }\right] \gamma +\frac{n}{%
\ell ^{2}\Gamma \left( M_{0},n\right) }\frac{1}{\gamma }\right\} ,  \label{9}
\end{equation}%
\ where we have defined

\begin{equation}
\mathbb{M}_{\nu m,s}^{\mp }=\mathbb{F}_{\nu m,s}^{\mp }+\frac{1}{\Gamma
^{2}\left( M_{0},n\right) }\mathbb{G}_{\nu m,s}^{\mp ,\alpha },  \label{10}
\end{equation}%
with%
\begin{eqnarray*}
\mathbb{F}_{\nu m,s}^{\mp } &=&\frac{\nu !}{(\nu +\left\vert m\right\vert
)!\ }\int_{0}^{\infty }dr\ r^{\left\vert m\right\vert -1/2}\exp (-r)\
\left\vert L_{\nu }^{\left\vert m\right\vert }\left( r\right) \right\vert
^{2}, \\
\ \ \ \ \mathbb{G}_{\nu m,s}^{\mp ,\alpha } &=&\frac{\nu !}{(\nu +\left\vert
m\right\vert )!\ }\int_{0}^{\infty }dr\ r^{\left\vert m\right\vert -1/2}\exp
(-r)\left\{
\begin{array}{lc}
r\ \left\vert L_{\nu +\alpha }^{\left\vert m\right\vert +1}\left( r\right)
\right\vert ^{2} & \alpha =\left\{
\begin{array}{cc}
0 & \text{if }m\geq 0\text{ and }s=+1 \\
-1 & \text{if }m<0\text{ and }s=-1\text{\ }%
\end{array}%
\right.  \\
n^{2}r^{-1}\ \left\vert L_{\nu +\alpha }^{\left\vert m\right\vert -1}\left(
r\right) \right\vert ^{2} & \alpha =\left\{
\begin{array}{cc}
0 & \text{if }m\geq 0\text{ and }s=-1 \\
+1 & \text{if }m<0\text{ and }s=+1\text{\ }%
\end{array}%
\right.
\end{array}%
\right. .
\end{eqnarray*}%
Minimizing Eq.~(\ref{9}) with respect to variational parameter $\gamma ,$ we
obtain

\begin{equation}
\gamma =\mp \left\{ \frac{n}{\ell ^{2}\Gamma \left( M_{0},n\right) }/\left[
\frac{2n}{\Gamma \left( M_{0},n\right) }-Z\alpha \mathbb{M}_{\nu m,s}^{\mp }%
\right] \right\} ^{1/2}.  \label{11}
\end{equation}%
where we restrict ourselves to the physical positive sign of the
variational parameter, due to the sign convention we employed in the
normalization \cite{griff,WGREINER}. Here, it should also be noted
that, in the literature, different sign conventions may be choosen,
however some of them may introduce artificial difficulties,
especially, in the massless limit ( Ref.~\onlinecite{griff} for a
detailed discussion). Therefore, by  labeling the electron and hole
states $\lambda =+1 $ and $\lambda =-1 $, respectively,  we obtain
an analytical result for the whole energy spectrum of the system
with Coulomb impurity potential as

\begin{equation}
\overline{E}_{n}^{\lambda }=\lambda \left( M_{0}^{2}+\frac{2n}{\ell ^{2}}%
\right) ^{-1/2}\left\{ M_{0}^{2}+\frac{2n}{\ell ^{2}}\left[ 1-\mathbb{H}%
\left( Z,M_{0}\right) \right] ^{1/2}\right\} ,  \label{12}
\end{equation}%
with

\begin{equation*}
\mathbb{H}\left( Z,M_{0}\right) =\lambda \frac{Z\alpha \mathbb{M}_{\nu
m,s}^{\mp }}{2\sqrt{2}}\frac{\ell }{n}\left( M_{0}+\sqrt{M_{0}^{2}+\frac{2n}{%
\ell ^{2}}}\right) ,
\end{equation*}%
for all values of $Z$ and $M_{0}$ if and only if $\mathbb{H}\left(
Z,M_{0}\right) \leq 1$, which implies the condition

\begin{equation}
\left( Z\alpha \right) _{cr}\equiv \overline{Z}_{cr}\left( M_{0},B\right) =%
\frac{2\sqrt{2}n/\ell }{\lambda \mathbb{M}_{\nu m,s}^{\mp }\left[ M_{0}+%
\sqrt{M_{0}^{2}+2n/\ell ^{2}}\right] }  \label{13}
\end{equation}%
should be satisfied in order that there might exist oscillator like discrete
energy spectrum. Otherwise, Eq.~(\ref{12}) becomes imaginary. In $\mathbb{H}%
\left( Z,M_{0}\right) =0$ limit, Eq.~(\ref{12}) becomes $\overline{E}%
_{n}^{\pm }=\pm \sqrt{M_{0}^{2}+2n/\ell ^{2}}$. Thus, we recover the usual
Landau picture of 2D Dirac equation with mass.

\begin{figure*}[b]
\centering
\includegraphics* [ height=8cm,width=5cm]{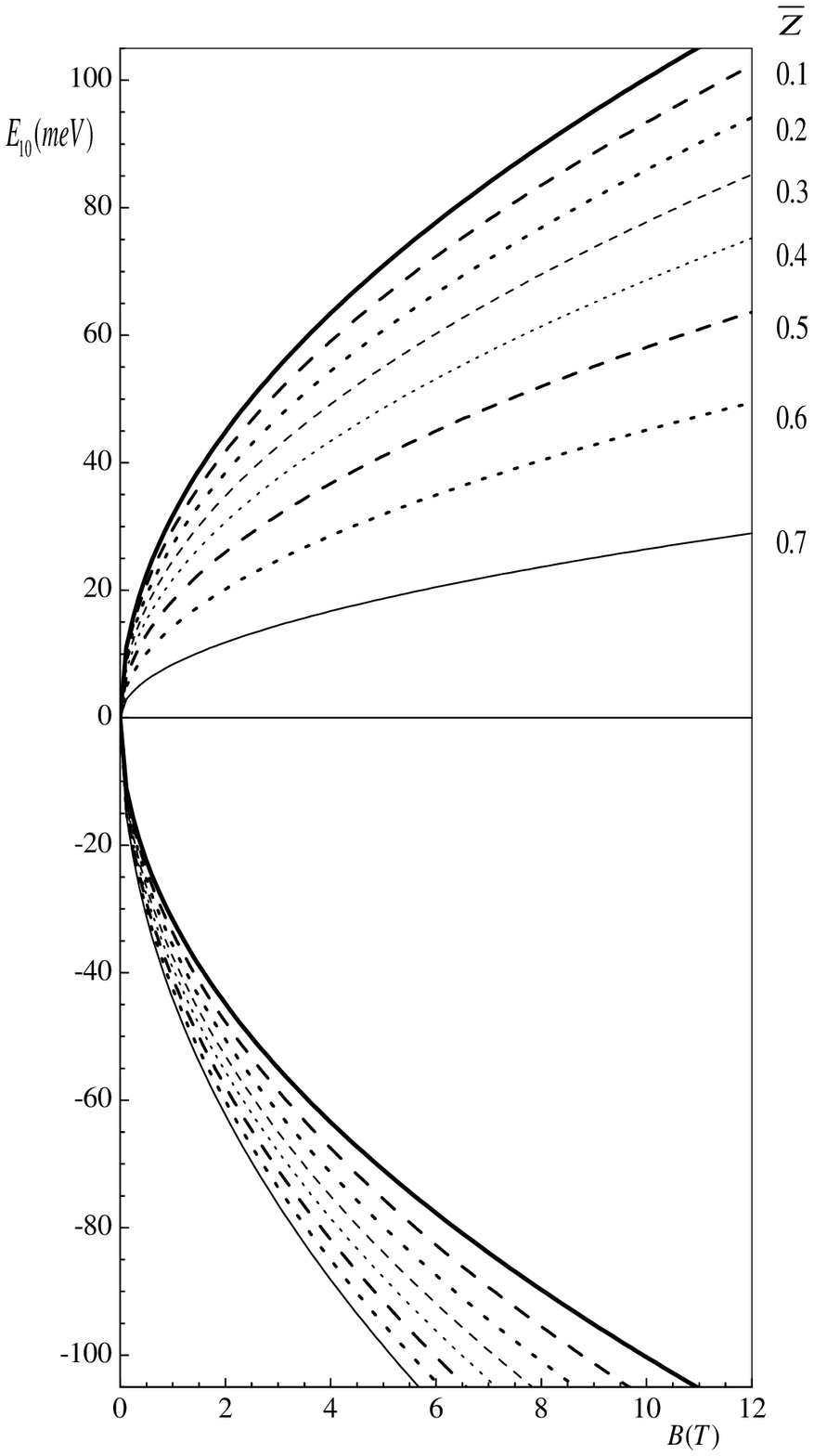} (a) %
\includegraphics* [ height=8cm,width=5cm]{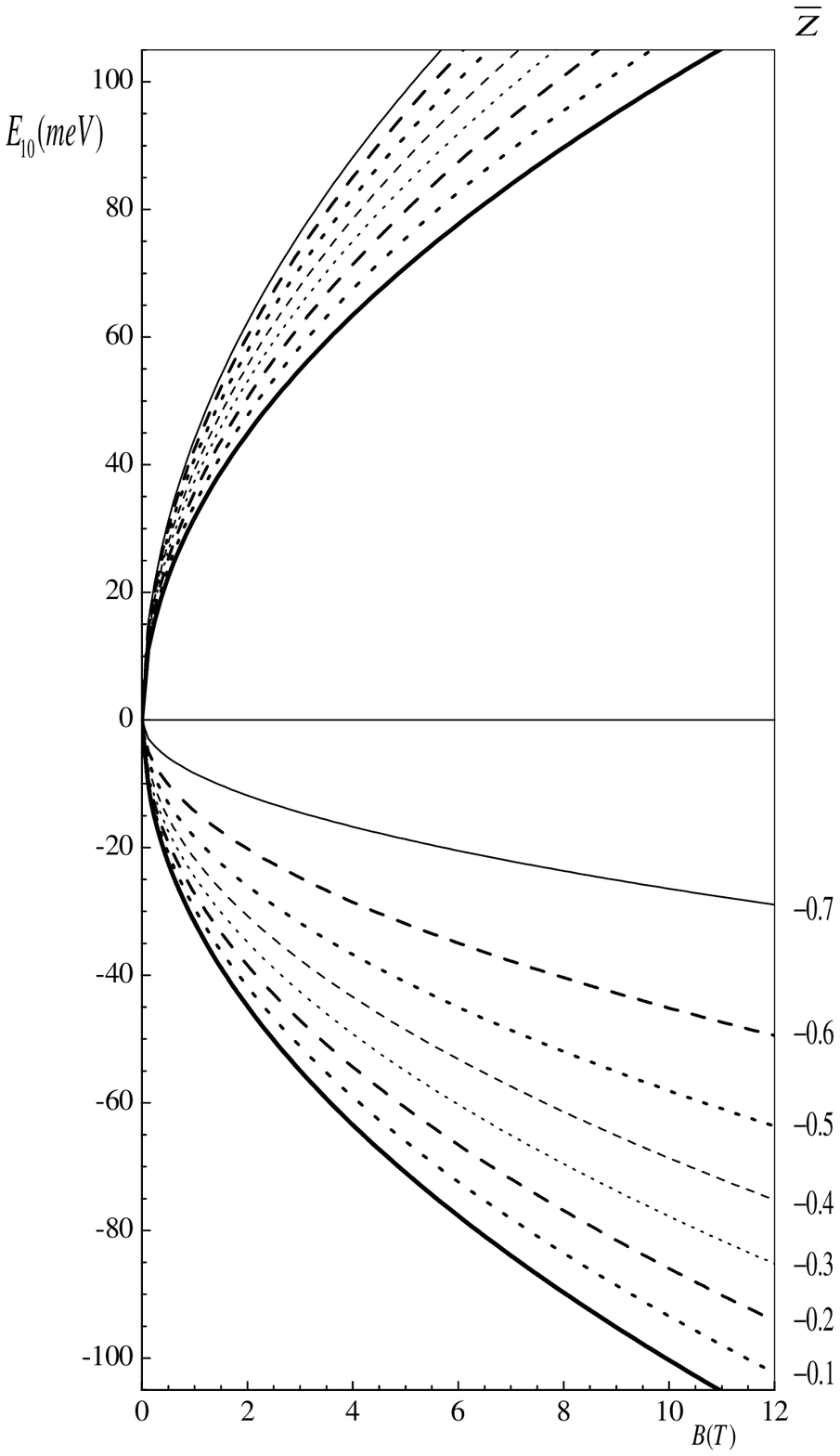} (b)
\caption{(a) Plots of the lowest (i.e., $n=1$ corresponding to the levels $%
\left( \protect\nu ,m,s\right) =\left( 0,1,-1\right) $, and $\left(
0,0,+1\right) $) Landau energy $\overline{E}_{1}^{(\pm )}$ of
gapless graphene electron and hole, calculated by
Eq.~(\protect\ref{12}) as a function of magnetic field for various
positive values of $\overline{Z}$. (b) Same as (a), but for negative
values of $\overline{Z}$.} \label{FIG1}
\end{figure*}

It should be noted that $\overline{E}_{n}^{(\pm )}$ given by Eq.~(\ref{12})
depends strongly on the sign of $\overline{Z}$, i.e., it is asymmetric with
respect to reverse of Z, due to the fact that, in the presence of a uniform
magnetic field, electron (hole) in a attractive impurity potential $(%
\overline{Z}>0,(\overline{Z}<0))$ does not have the same energies as
electrons (holes) in a repulsive impurity potential $(\overline{Z}<0,(%
\overline{Z}>0))$. This is apparent from FIG.~\ref{FIG1} (a) and (b)
where the evolution of the ground state energy,
$\overline{E}_{1}^{(\pm )}$ of massless graphene electron(hole) in
the magnetic field is shown for (a)
various values of $\overline{Z}>0$, and (b) for various values of $\overline{%
Z}<0$. As is seen from the figure, when $\overline{Z}\ $  approaches
to $\pm \overline{Z}_{cr}$, $ \overline{E}_{1}^{(\pm )}$ is
dramatically reduced, and it vanishes at $\overline{Z}=\pm
\overline{Z}_{cr}$. It can be easily
seen from Eq.~(\ref{13}) that, in the case of massless graphene, i.e., $%
M_{0}=0,$ this critical value of $\overline{Z}$, i.e., $\overline{Z}_{cr}$
is given by

\begin{equation}
\overline{Z}_{cr}\left( 0,B\right) =\pm \frac{2}{\sqrt{n}}\frac{1}{\mathbb{M}%
_{\nu m,s}^{\mp }\mid _{M_{0}=0}}  \label{14}
\end{equation}%
which is independent of $\ell $ , i.e., of $B$, and it leads to $\overline{Z}%
_{cr}=4/3\sqrt{\pi }=0.752$ for the ground-state of the system, i.e., for $%
n=1$ corresponding to $\left( \nu ,m,s\right) =\left( 0,1,-1\right) $, and $%
\left( 0,0,+1\right) $. In fact, it is  well-known  that, in the
absence of magnetic field, this critical value is equal to
$0.5$\cite{valeri,shytov2} for the massive case. This difference
arises from the use of oscillator states as trial states in our
problem to obtain the dependence of critical value of
$\overline{Z}$ on magnetic field. Therefore, it is obvious that
our analytical results are valid for high magnetic field regime.

\begin{figure}[h]
\centering
\includegraphics[ height=10cm,width=5cm]{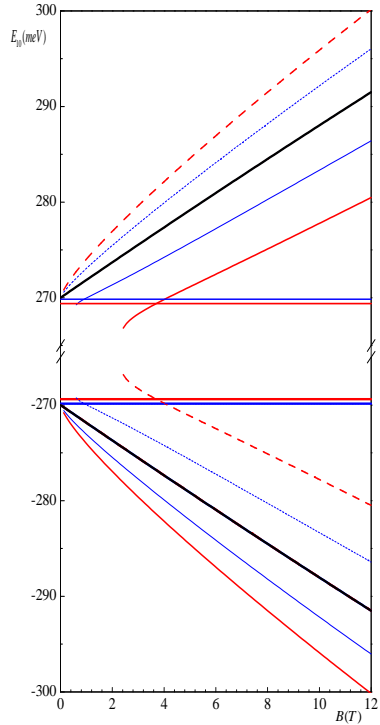}
\caption{(Color online) Dependence of the lowest Landau Level of
the gapped graphene on magnetic field, calculated by
Eq.~(\protect\ref{12}). While the thin and thick dashed curves
mark $\overline{Z}=-0.05 $ and $\overline{Z}=-0.1$, respectively,
the thin and thick solid lines mark $\overline{Z}=0.05$ and
$\overline{Z}=0.1$, respectively. The thick bold lines corresponds
unperturbed $\overline{E}_{1}^{\pm }=\pm
\protect\sqrt{M_{0}^{2}+\left( 2/\ell ^{2}\right) }\ $ energy
levels, i.e., in the absence of impurity, graphene electron and
hole ground-state energies, respectively. The horizontal solid
lines refer to the associated impurity eigenvalues
in the absence of magnetic field, i.e., Eq.~(\protect\ref{15}) with $n=1,m=0$%
.} \label{FIG2}
\end{figure}

In the presence of a gap/"mass" term, as shown in FIG.~\ref{FIG2},
the picture is quite different depending on the values of
$\overline{Z}$, compared with those of found in FIG.~\ref{FIG1} (a)
and (b). In this figure, we use a gap of value $\sim
0.26\,\mathrm{eV}$ which is reported in single layer graphene due to
$\mathrm{SiC}$ substrate by Zhou et al\cite{zhou}, and we choose
$M=0.1t$
which corresponds to $270\,\mathrm{meV}$ with hopping energy $t=2.7\,\mathrm{eV}$%
. Curves for
\begin{equation}
\overline{E}_{n}\left( M_{0},\overline{Z}\right) =M_{0}sgn\left( \overline{Z}%
\right) \left\{ 1+\frac{\overline{Z}^{2}}{\left[ n+\sqrt{\left( m+1/2\right)
^{2}-\overline{Z}^{2}}\right] ^{2}}\right\} ^{-1/2}  \label{15}
\end{equation}%
without magnetic field but with impurity\cite{WGREINER}, \ and curves for $%
\overline{E}_{n}^{\mp }\left( M_{0},B\right) =\mp
\sqrt{M_{0}^{2}+\left( 2n/\ell ^{2}\right) }$ with magnetic field
but without impurity are also plotted in the same figure for
comparison purposes. Here, $sgn$ is the sign function. We represent
the former group, i.e., electron and hole energy bands, by
horizontal solid lines, while the bold straight lines are used for
the later one. We first notice that, with inclusion of gap/"mass"
term, the behavior of the ground-state energy changes from a square
root of $B$ dependence to a linear dependence on $B$. As $B$ is
decreased, the curves for electron energies with $\overline{Z}=-0.1$
and $-0.2$, and the curves for the hole
energies with $\overline{Z}=0.1$ and $\ 0.2$, all they approach $\overline{E}%
_{n}\left( M_{0},\overline{Z}\right) $ indicated by horizontal lines which
correspond to impurity energy in the absence of magnetic field. It should
also be noted that, while the curve for gapped-graphene electron (hole)
energy with $\overline{Z}=0.05$ ($-0.05$) spreads very little through the
gap,  the curve for gapped-graphene electron (hole) energy with $\overline{Z}%
=0.2$ ($-0.1$) lies significantly below the gap.

\begin{figure*}[h]
\includegraphics* [height=8cm,width=5cm]{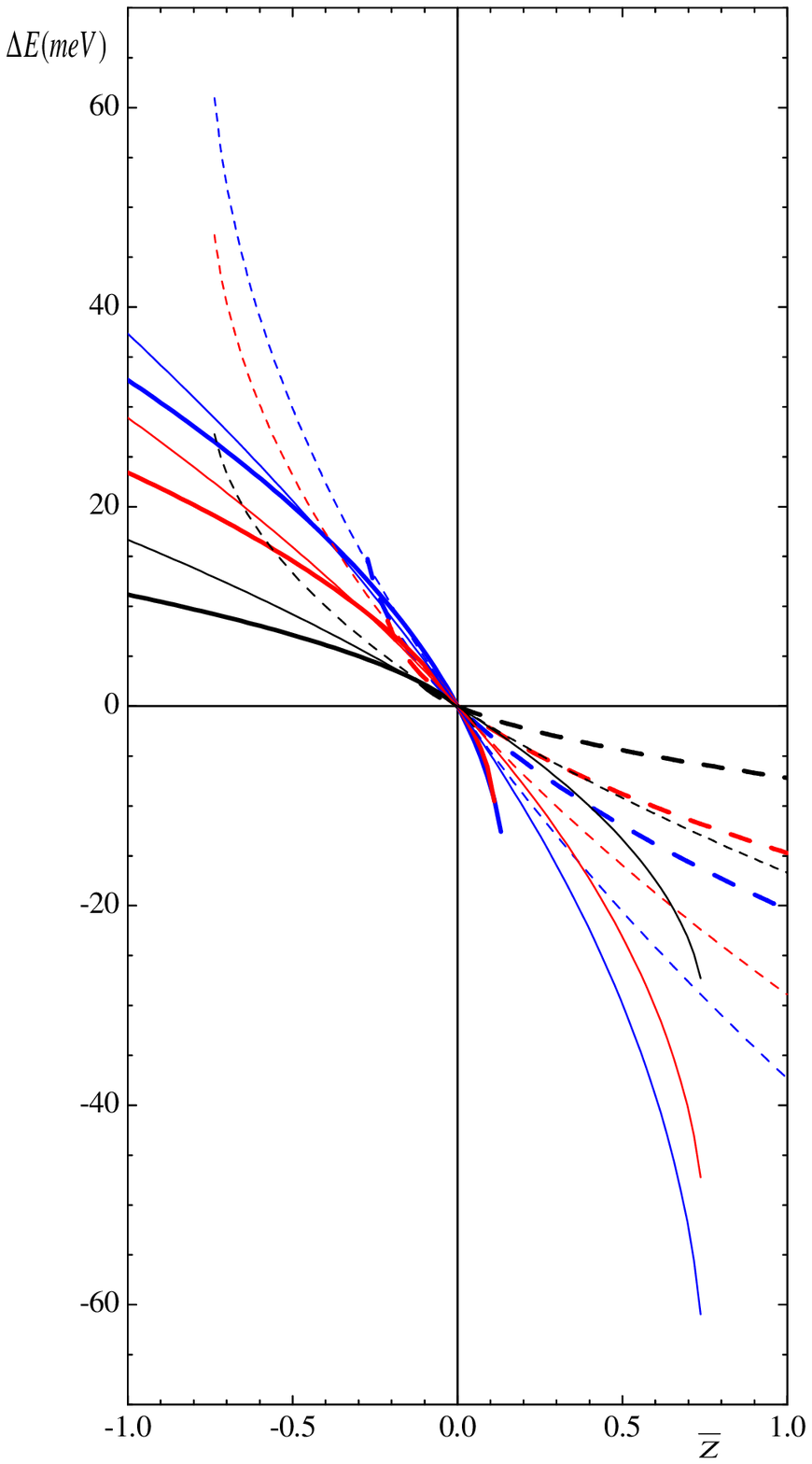} (a) %
\includegraphics*  [ height=8cm,width=5cm]{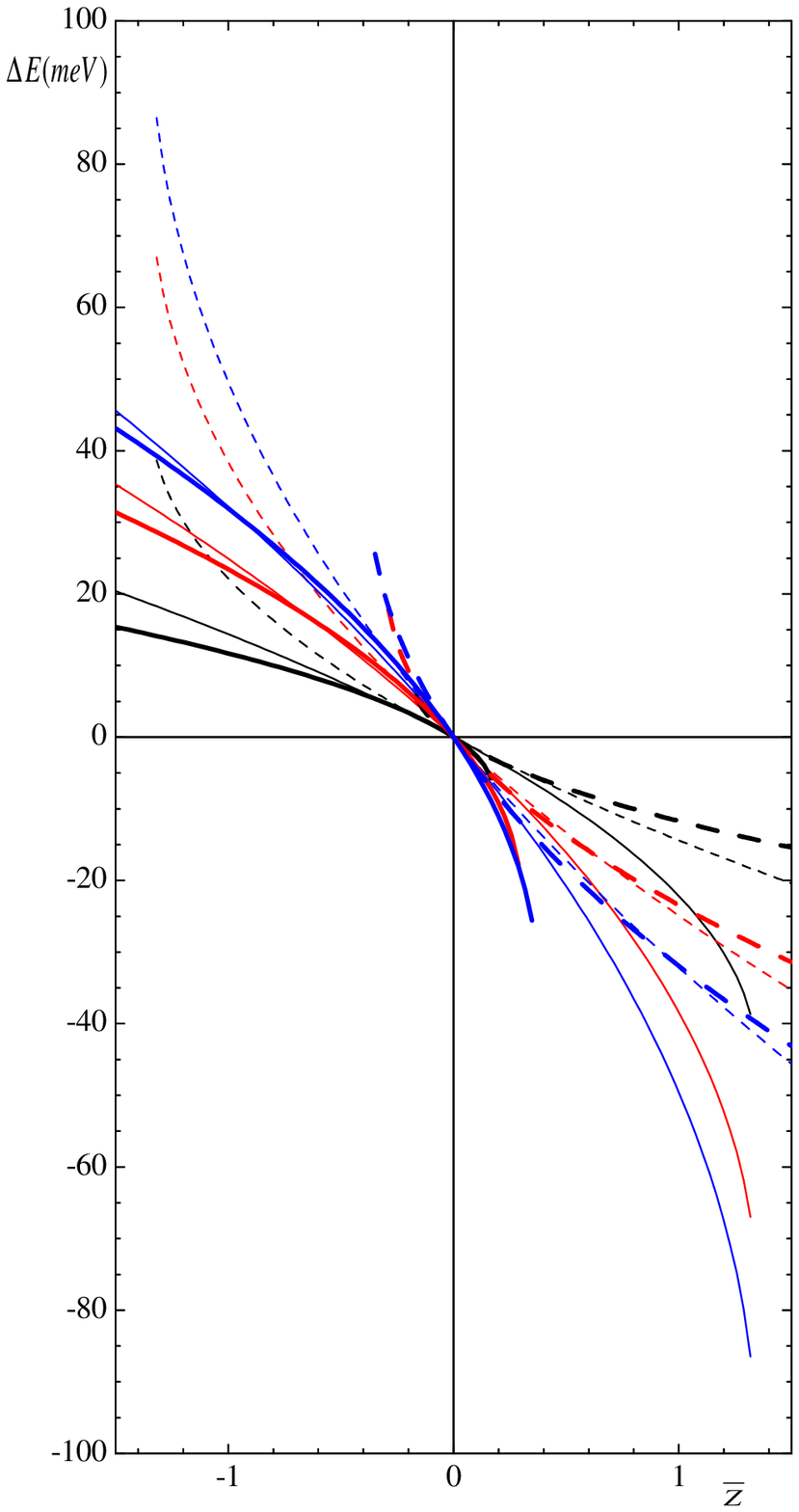} (b) %
\includegraphics* [ height=8cm,width=5cm]{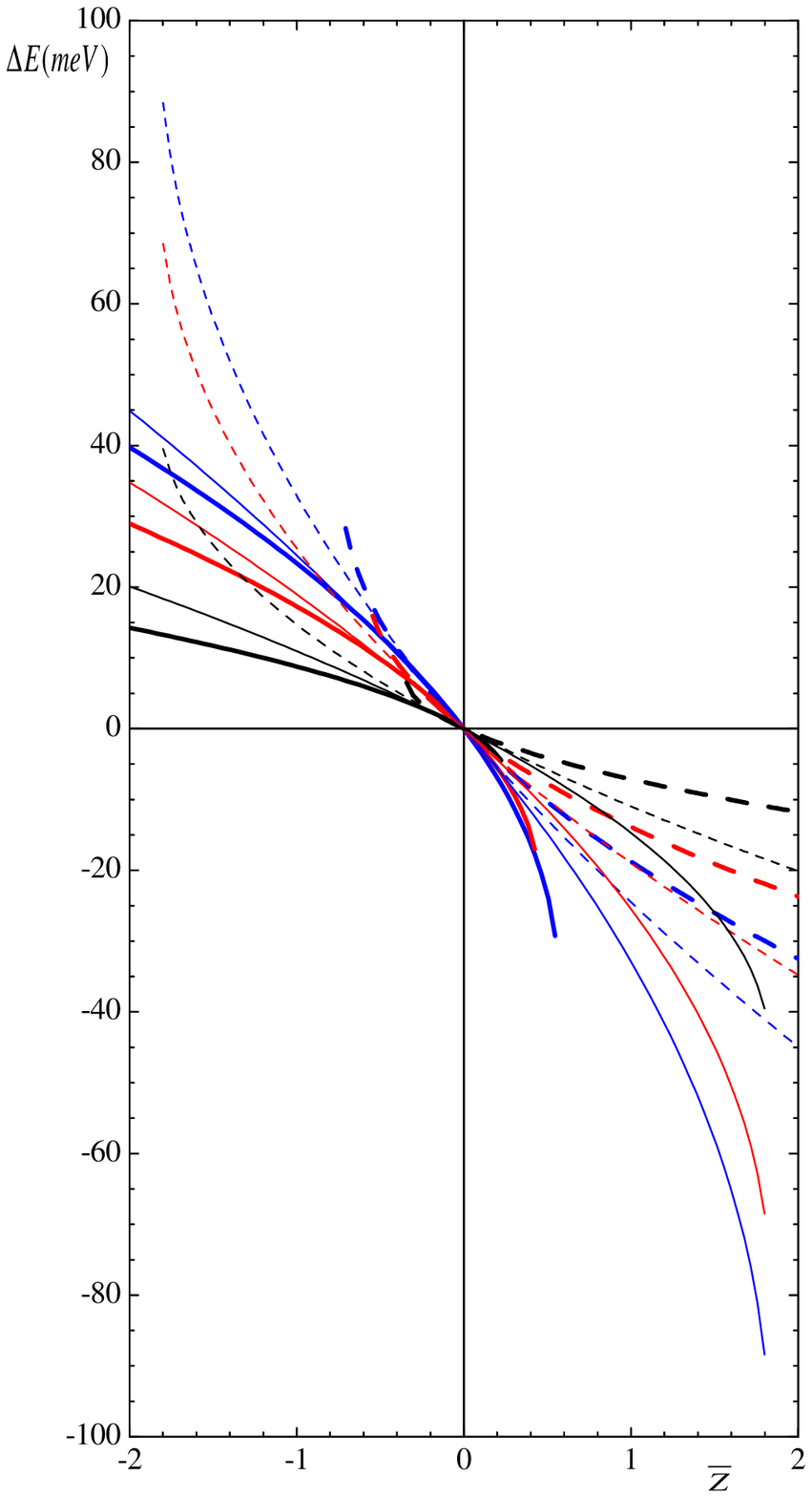} (c)
\caption{(Color online) Impurity binding energies of gapless
graphene electron (thin solid lines) and gapless graphene hole (thin
dashed lines) as a function of $\overline{Z}$ for the three
low-lying Landau levels, i.e., (a) for $(10)$, (b) for $(20)$, (c)
for $(21)$, for three values of magnetic field, $B=1$ (black)$,3$
(red) and $5$T (blue). Binding energies of gapped graphene electron
and hole states are also shown in the same graphs but by thick
counterparts for $\overline{M}_{0}=M_{0}/t=0.1$.} \label{FIG3}
\end{figure*}

To understand this picture better, it is necessary to study the
impurity binding energy, which is a measure of how much of
$\overline{E}_{n}^{\pm }$ consists of \ the charged Coulomb
impurity. In other words, it is defined as the energy difference
between the energy of graphene (massless or massive) without
impurity and the energy of graphene with impurity, i.e., $\Delta
\overline{E}_{n}^{\pm BE}=\pm \sqrt{M_{0}^{2}+\left( 2n/\ell
^{2}\right) }-$ $\overline{E}_{n}^{\pm }$. In FIG.~\ref{FIG3}(a-c),
impurity binding energies for low- lying gapless and gapped-graphene
states are given for three different values of magnetic field
strength. From these figures, we see that enhancement in magnetic
field strength leads to more binding. Moreover, in the absence of
gap term, binding energies for electrons (thin lines) and holes
(thin dashed lines) have the same magnitude, i.e., they are all
equal in magnitude under $\overline{Z}\rightarrow -\overline{Z}$.

\begin{figure*}[h]
\includegraphics* [ height=8cm,width=6cm]{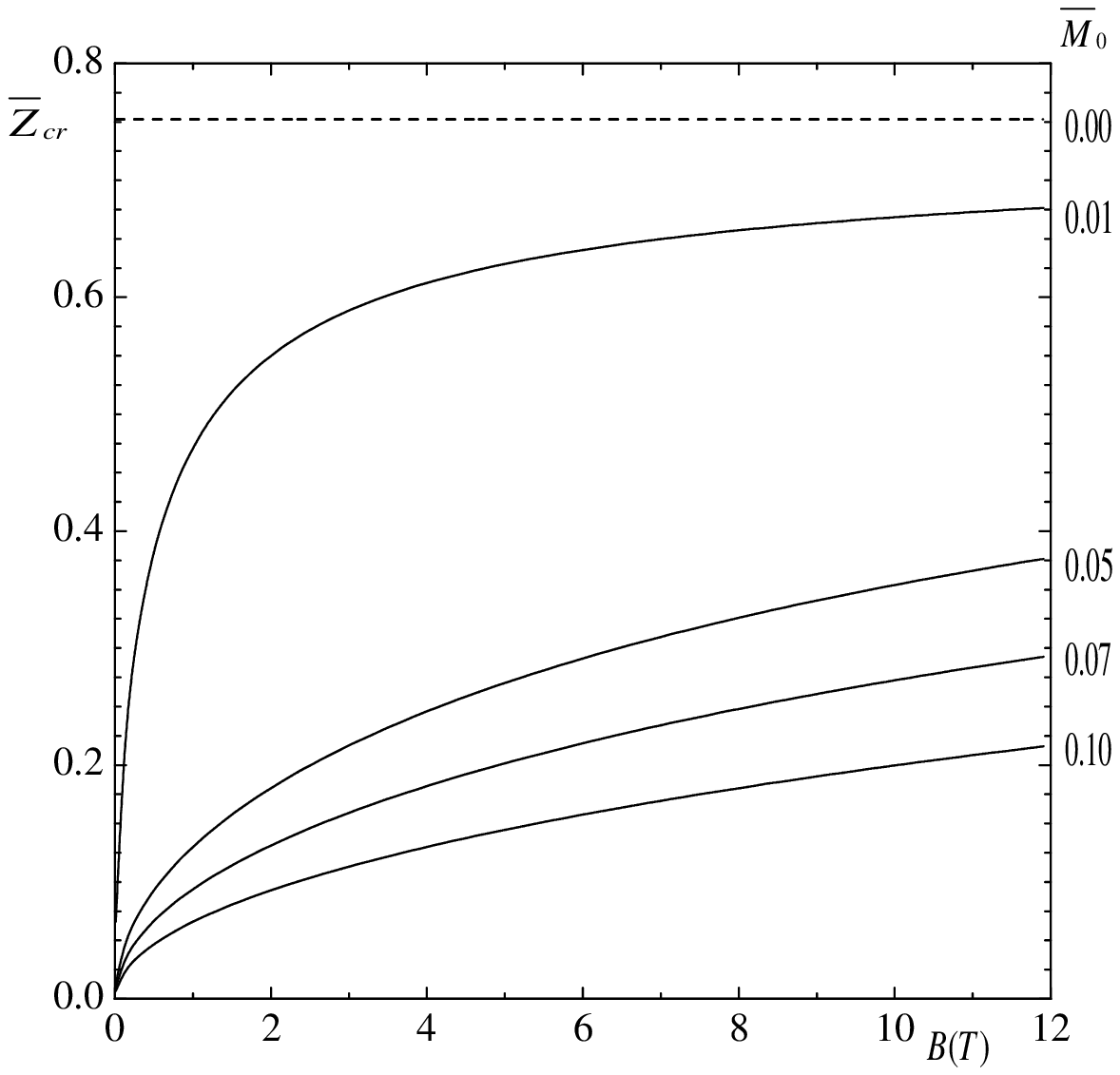} (a) %
\includegraphics*[ height=8cm,width=6cm]{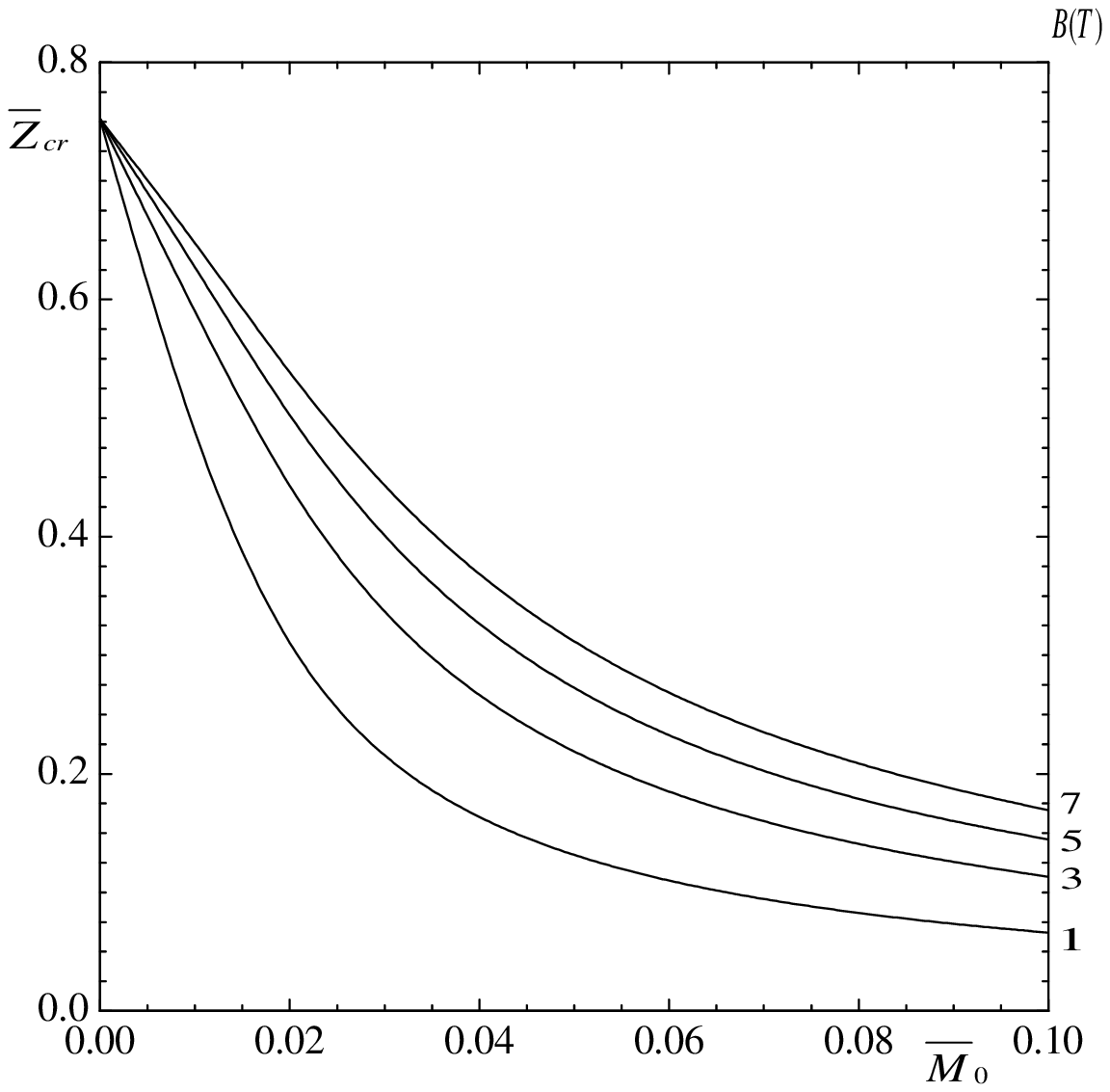} (b)
\caption{(a)The critical value of $\overline{Z}_{cr}$ given by
Eq.~(\protect \ref{13}) as a function of magnetic field for four
different gap values, and (b) as a function of gap parameter, but
for four different magnetic field values. } \label{FIG4}
\end{figure*}

To investigate the influence of gap/"mass" term in detail, we have
also examined the dependence of $\overline{Z}_{cr}$ on both magnetic
field strength\ $B$ and gap/mass term $M_{0}$, in FIG.~\ref{FIG4}(a)
and (b), respectively. The curves of the figure corresponds to
$\overline{E}_{1}^{+}$ case only, since tendencies of changing of
$\overline{E}_{1}^{-}$ with $B$ and $M_{0}$ are similar to those of
$\overline{E}_{1}^{+}$. For a given
value of $M_{0}$, as $B$ is increased, the minimum value of $\overline{Z}%
_{cr}$ curves shifts to larger values. As expected, all figures in FIG.~\ref%
{FIG4}(a) and (b) illustrate that, as $\overline{M}_{0}$ approaches
to zero, threshold value of $\overline{Z}_{cr}$ is given by $0.752$
as is calculated above. By switching a gap, $\overline{Z}_{cr}$ is
drastically reduced compared to that found in the massless limit.

\begin{figure*}[h]
\includegraphics*[ height=8cm,width=6cm]{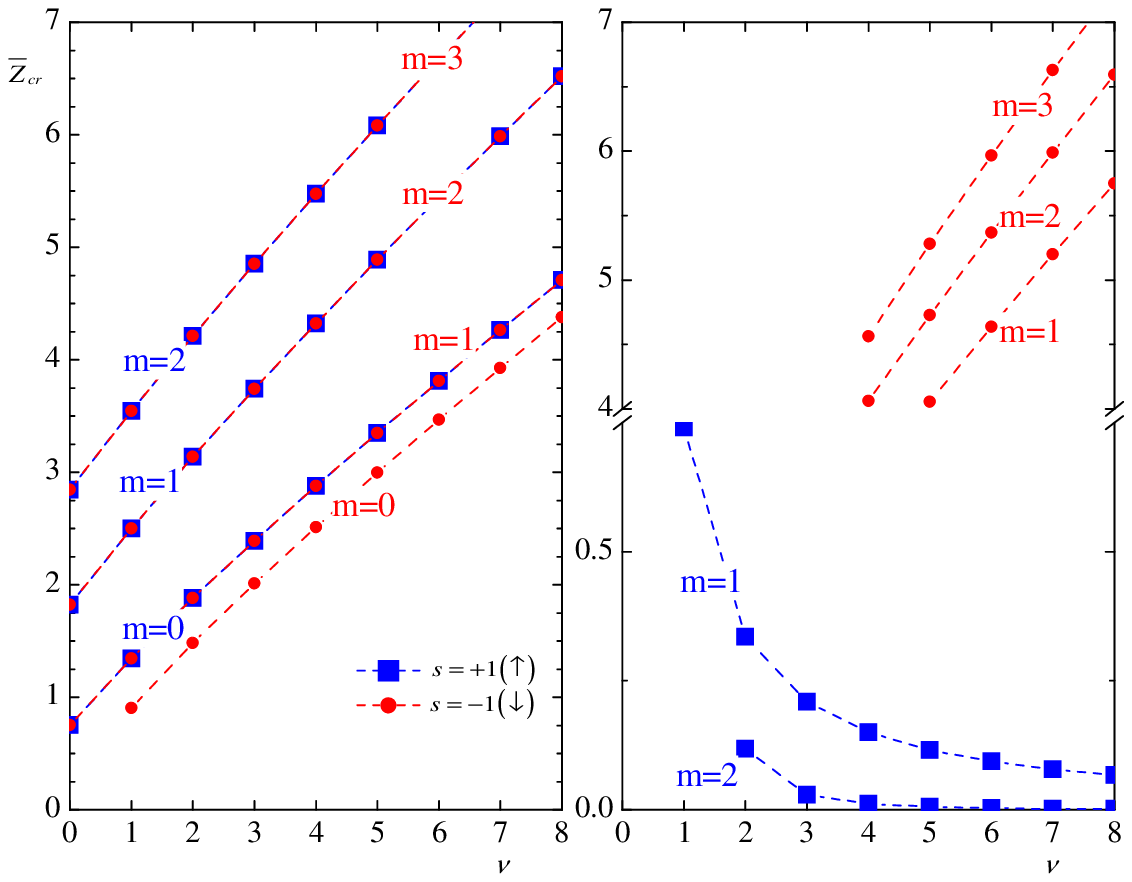} (a) %
\includegraphics* [height=8cm,width=6cm]{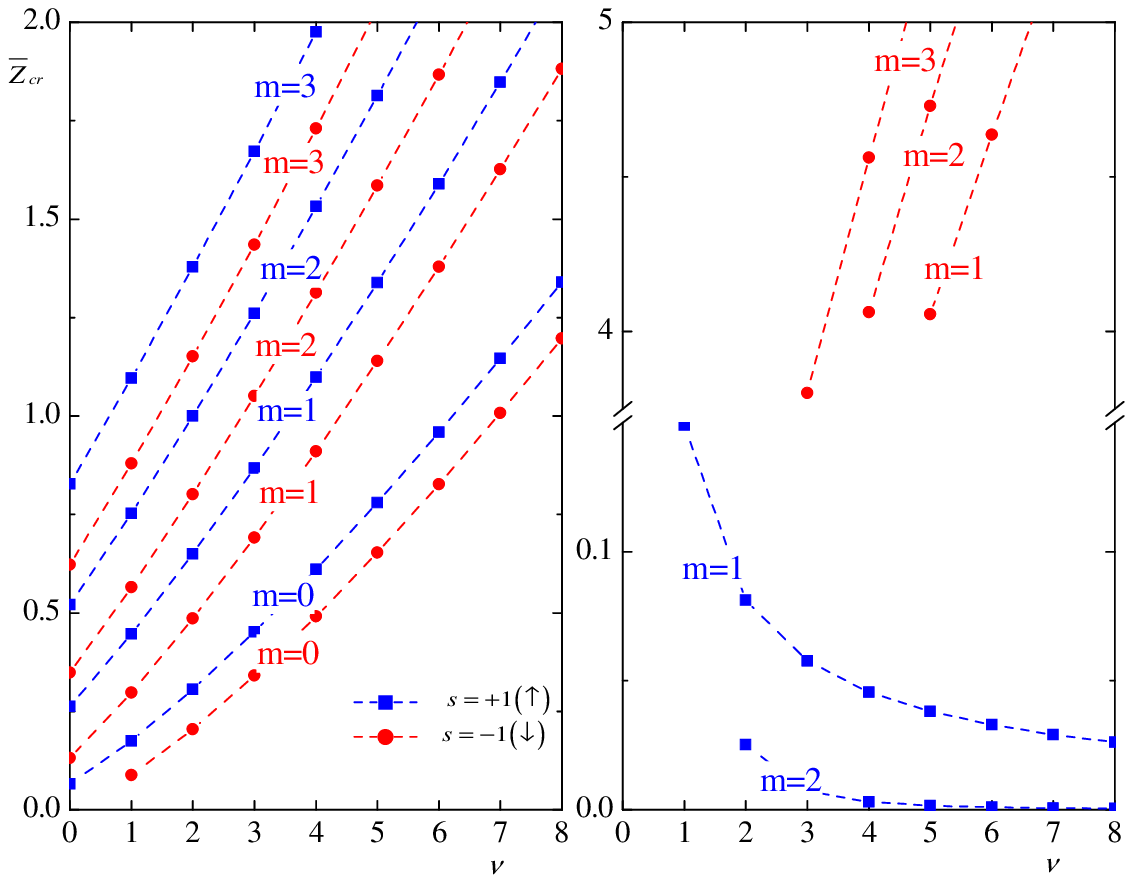} (b)
\caption{(Color online) Dependence of the critical value of Coulomb
coupling strength, i.e., $\overline{Z}_{cr}$ on quantum numbers
$(\protect\nu m)$ for (a) without gap, $\overline{M}_{0}=0$ (b) with
gap, $\overline{M}_{0}=0,1$, respectively. The dashed lines are
drawn to guide the eye. In these figures, left panels corresponds to
$m\geq 0$, while the right ones refer to $m<0$.} \label{FIG5}
\end{figure*}

In FIG.~\ref{FIG5}(a) and (b) for comparison purposes, we plotted
the evolution of $\overline{Z}_{cr}$ as a function of $(\nu m)$ for
massless and massive cases, respectively. The energy levels are
dependent on the
principal quantum number $n$, and they are degenerate with respect to $m$, $%
-m$ and $s$. By comparing the left and right panels of
FIG.~\ref{FIG5}(a), we see that the degeneracy of energy levels with
respect to $m$, $-m$ \ are partially removed by the impurity.
However, inclusion of a gap/"mass" term (
FIG.~\ref{FIG5}(b)) \ splits the degeneracy of these levels with respect to $m$%
, $-m$ and $s$. Additionally, by switching a gap/"mass" term, we see
that \ critical values of $\overline{Z}$ are (i) smaller than those
presented for the massless limit, and (ii) they diminish when
increases for $\left( \nu ,m<0,+\right) $ while $\overline{Z}_{cr}$
for $\left( \nu ,m<0,-\right) $ states increases. These are due to
the fact that, while the massless Dirac-Weyl Hamiltonian with
external fields commutes with helicity operator, i.e., it is
invariant under the chiral transformation, the mass term is not,
therefore it breaks the chiral symmetry. For the physical origin of
this axial anomaly, we refer to the book of Huang \cite{Huang} where
a detailed discussion of nonexistence of conserved axial vector
current is given in a
whole chapter, and we also refer to the paper of Lee \textit{et al} \cite%
{Lee} wherein they concluded that first chiral states does not decouple,
even in the massless limit, and second massless spinor electrodynamics is a
pathological theory and therefore it should always be considered to be the
limit of a massive theory.

In conclusion, we have investigated the effect of charged Coulomb
impurity on the Landau level spectrum of both gapless and
gapped-graphene near the Fermi point, by using their 2D continuum
massless and massive Dirac descriptions, respectively, in the
framework of a variational procedure. Since, we know that the
success of variational treatment strongly depends on the choice of
trial wave functions together with the set of parameters included,
we suggest the basis sets of unperturbed Hamiltonian as trial wave
functions with one parameter so as to reproduce the exact analytical
results in the absence of perturbation. Indeed, to test the results
of our variational calculation, it is enough to look at the change
of variational parameter as a function of impurity strength $Z$. In
the limit $Z\rightarrow 0$, the variational parameter becomes
$\gamma=1/\ell$. This is an indication of the adequacy of the trial
wave function, since it yields the exact results in the absence of
impurity, as expected. Furthermore, in the absence of impurity the
variational parameter $\gamma=1/\ell$ is the inverse order of
magnitude of magnetic confinement length. When impurity is switched
on, renormalization of the magnetic confinement length should be
taken into account so as to adapt itself to the response of the
presence of impurity, for which the energy develops a new minimum.
This is achieved by adjusting the variational envelope wave function
with a variational parameter $\gamma$ until the energy is minimized.
Thus, the inverse of $\gamma$ may be interpreted as the amount of
effective displacement of the magnetic confinement length caused by
the impurity. Although the results of variational calculation we use
give good agreement with the exact ones in the absence of impurity,
and they also yield exact analytical results for the ground- and
exited-states in the presence of impurity, such a description may
not be accurate for the Landau bands with high indices in the case
of high magnetic fields. In this case, adjacent Landau bands may
overlap. As a consequence, a more sophisticated trial wave function
composed of linear superpositions of trial wave functions we use
with more than one variational parameter may be needed for a
numerical approach. \cite{kandemir}

In summary, by employing a variational procedure based on a choice
of trial wave functions as the basis sets of Landau levels, we
obtained analytical results for both energy eigenvalues and the
critical Coulomb coupling strength as functions of both magnetic
field strength and gap parameter for a graphene electron subjected
to a homogeneous magnetic field in the presence of a charged Coulomb
impurity. The analytical results that we obtained here show that (i)
the critical impurity strength is independent of the magnetic field
strength in the massless limit, and (ii) it drastically reduces when
a gap/"mass" term is switched on. However, enhancement in magnetic
field strength leads to increase in critical Coulomb coupling
strength.

\begin{acknowledgments}
The authors thank Professor T. Altanhan for valuable discussions, and for a
critically reading of the manuscript.
\end{acknowledgments}

\end{document}